# Spatially multiplexed dielectric tensor tomography

JUHEON LEE,[1,2] SEUNGWOO SHIN,[3] HERVE HUGONNET,[1,2] AND YONGKEUN PARK[1,2,4*]

[1]Department of Physics, Korea Advanced Institute of Science and Technology (KAIST), Daejeon 34141, Republic of Korea.
[2]KAIST Institute for Health Science and Technology, KAIST, Daejeon 34141, Republic of Korea.
[3]Department of Physics, University of California at Santa Barbara, Santa Barbara, CA 93106, USA.
[4]Tomocube Inc., Daejeon 34109, Republic of Korea.
*Corresponding author: yk.park@kaist.ac.kr

**Dielectric tensor tomography (DTT) enables the reconstruction of three-dimensional (3D) dielectric tensors, which provides a physical measure of 3D optical anisotropy. Herein, we present a cost-effective and robust method for DTT measurement using a multiplexing approach. By exploiting two orthogonally polarized reference beams with different angles in an off-axis interferometer, two polarization-sensitive interferograms were overlapped and recorded using a single camera. Then, the two multiplexed interferograms were separated in the Fourier domain, enabling the reconstruction of polarization-sensitive fields. Finally, by measuring the polarization-sensitive fields for various illumination angles, 3D dielectric tensor tomograms were reconstructed. The proposed method was experimentally demonstrated by reconstructing the 3D dielectric tensors of various liquid crystal particles with radial and bipolar orientational configurations.**

The anisotropic nature of a material, light-matter interactions that depend on the polarization of light, is its distinctive optical property. Anisotropy provides information about the speed of light in the media, as well as the crystalline structure orientation through the measurement of principal refractive indices (RI) and directors. This information is exploited in materials science, biology, and industrial applications [1-3]. Conventionally, anisotropy can only be partially exploited using polarized light microscopy [4, 5]. By measuring two-dimensional (2D) polarization-sensitive transmission intensity images, birefringence information is deduced qualitatively in 2D, and out-of-plane directional information is lost [Fig. 1(a)]. 2D polarization-sensitive quantitative phase imaging (QPI) techniques have been developed to measure quantitative phase delays of birefringence samples [6-8].

3D QPI techniques, such as optical diffraction tomography (ODT) [9-11], measure the 3D RI distribution. Although the conventional use of ODT is limited to isotropic samples [Fig. 1(b)], polarization-sensitive ODT, which reconstructs the 3D RI distribution for different illuminations and analyzer polarizations [12, 13] has also been developed. However, anisotropy information is provided without exact 3D anisotropic directions because one dimension is neglected to deal with the lack of degrees of freedom. Recently, the underdetermined problem was solved using dielectric tensor tomography (DTT) [14] [Fig. 1(c)]. This principle exploits the Fourier differentiation theorem to obtain additional degrees of freedom. This approach enables the direct acquisition of 3D dielectric tensors, a physical quantity describing the light-matter interaction of anisotropic structures with various principal RIs and orientations of the optical axes. To reconstruct the dielectric tensor tomogram, DTT measures polarization-sensitive fields at various illumination angles and vectorizes these fields in the Cartesian coordinate system to solve the vector wave equation.

Polarization-sensitive measurements are required to retrieve and vectorize a diffracted field from an anisotropic structure. DTT employs a pair of cameras with orthogonal polarizers to simultaneously record two holograms at various illumination angles [14]. Although this method enables the retrieval of polarization-sensitive fields with high speed, accurate image registration between the recorded images is required for vectorization, otherwise position disagreement deteriorates the reconstruction quality. Instead of using two cameras, polarization-sensitive fields can be retrieved using a camera with rotations of a polarizer and an analyzer [15]. Although this method is intuitive, the mechanical movements and the requirements for multiple measurements result in disadvantages, including instability and slow speed. An alternative method using one camera and copies of the sample beam can measure polarization-sensitive fields in a single-shot [16, 17]. However, the setup is bulky, and because copies of the sample beam are generated at separate positions of the camera sensor, the effective field of view is reduced. Using holographic multiplexing is another approach to measure polarization-sensitive fields using a camera without field-of-view reduction [7, 18, 19]. However, this method has not been exploited for the DTT principle, which requires vectorization of the measured polarization-sensitive fields.

In this study, for robust and precise DTT measurement, we present multiplexed dielectric tensor tomography (mDTT), which reconstructs 3D dielectric tensors using only one camera and holographic multiplexing without field-image registration. Compared to the original DTT setup, which exploits a pair of cameras and polarizers, mDTT uses only one camera and two polarized reference beams. Each reference beam interferes with the sample beam to form two interferograms on the camera. Two polarization-sensitive fields with different fringe directions can be retrieved from these interferograms using Fourier space filtering [18-20]. Because the retrieved fields are obtained without polarizers in front of the camera, field vectorization is conducted by considering the polarization states of the reference beams. In this way, vector fields are obtained while scanning the illumination angle, enabling tomographic reconstruction of the dielectric tensor [14] without any image registration.

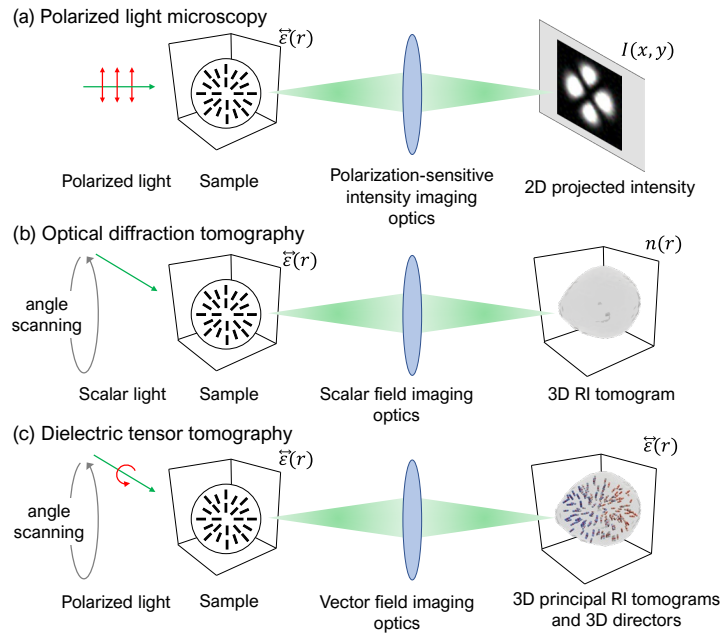

Fig. 1. Anisotropy imaging methods. (a) Polarized light microscopy. Anisotropy is qualitatively deduced from an intensity image. (b) ODT. Only the 3D RI distribution is reconstructed corresponding to the illumination polarization. (c) DTT. This method reconstructs the full anisotropy: the 3D principal RI distribution and its directions.

To demonstrate our method, we built a setup based on an off-axis Mach–Zehnder interferometer equipped with a diode-pumped solid-state laser (532 nm, Cobalt AB) [Fig. 2(a)]. The laser beam was split into a sample and reference arm using a polarizing beam splitter. To control the illumination angle, a digital micromirror device (DLi4130, Digital Light Innovations) was used [21, 22], followed by a liquid crystal retarder (LCC1223-A, Thorlabs) to control the polarization states. The polarized beam illuminated the sample through a condenser lens (UPLSAPO60XW, Olympus) and the light diffracted by the sample was collected using an objective lens (UPLXAPO60XO, Olympus). A Wollaston prism was used in the reference arm to generate two reference beams with orthogonal polarizations. To separate the beams equally, a half-wave plate rotated the polarization direction before the Wollaston prism. After the prism, a 4-$f$ imaging system was used to control the separation angle between the two reference beams. The sample beam and two reference beams were finally combined using a beam splitter before being detected using a camera (Lt425R, Lumenera) [Fig. 2(b)].

Because beams with perpendicular polarization do not interfere, the use of two reference beams with orthogonal polarization forms two interferograms multiplexed as a distinctive check pattern on the camera [Fig. 3(a)]. Each interferogram contains information regarding the polarization component of the sample beam parallel to the polarization of the relevant reference beam [18]. The angles of the reference beams are chosen such that neither of holograms overlap in the Fourier plane [Fig. 3(b)]. Therefore, the phase and intensity of each component can be retrieved independently [Fig. 3(c)] [18]. Detailed algorithms on field retrieval, phase retrieval, and tomographic reconstruction can be found elsewhere [23-25].

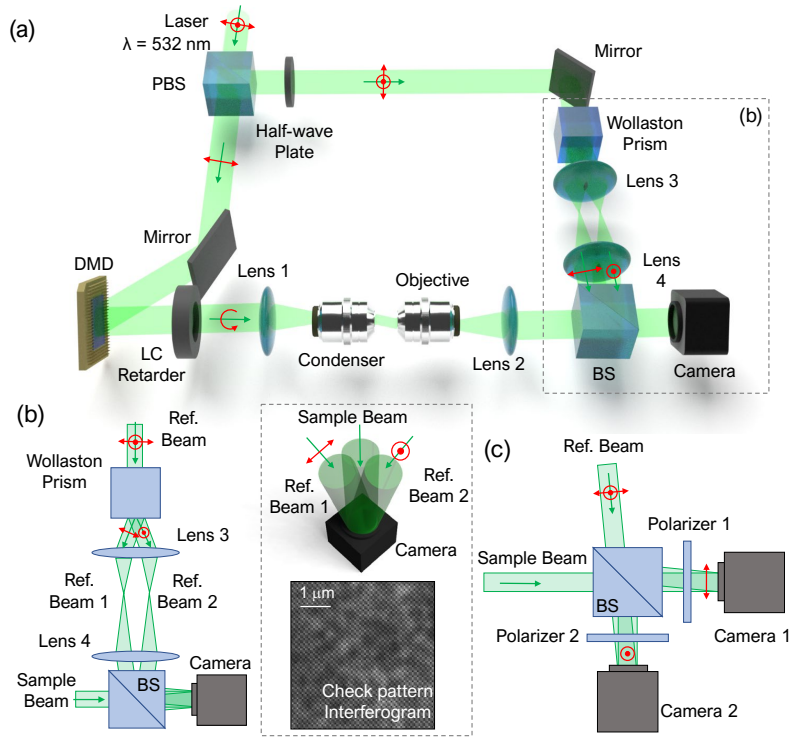

Fig. 2. (a) Multiplexed DTT setup. The setup is based on a Mach–Zehnder interferometry with DMD illumination. (b) Zoomed in image of the dashed box in (a). The reference beam is split into two polarization-orthogonal beams. Each beam interferes with a sample beam at different angles, making a multiplexed check pattern interferogram. (c) Counterpart of the dashed box in (a) for a conventional DTT setup. Two cameras are employed to capture two interferograms corresponding to orthogonal polarizations.

For DTT, the retrieved fields are vectorized in the Cartesian coordinate system [Fig. 3(d)]. The retrieved field can be expressed mathematically as the element wise dot product of the field with the reference beam. The fields $E_1$ and $E_2$ obtained using Fourier filtering,

$$E_1 = \overrightarrow{E_{cam}} \cdot \overrightarrow{R_1},\ E_2 = \overrightarrow{E_{cam}} \cdot \overrightarrow{R_2},\quad (1)$$

where $\overrightarrow{E_{cam}}$ denotes the unknown field at the camera plane and $\overrightarrow{R_1}$ and $\overrightarrow{R_2}$ represent the reference beams. Note that the polarization is a 3D vector, and to find the three components of the vector, the orthogonal relation between the polarization direction and wave vector should be used. However, because $\overrightarrow{E_{cam}}$ consists of various wave vectors, we make use of plane-wave decomposition. Any field can then be expressed as

$$\vec{E}(\vec{x}) = \iint_{k_x k_y} \widetilde{\vec{E_s}}(k_x,k_y)\vec{s}_{k_x,k_y} e^{i\vec{k}\vec{x}} + \widetilde{\vec{E_p}}(k_x,k_y)\vec{p}_{k_x,k_y} e^{i\vec{k}\vec{x}},\quad (2)$$

where the wave vector is expressed as

$$\vec{k} = (k_x, k_y, k_z)^T,\ k_z = \sqrt{k^2 - k_x^2 - k_y^2},\quad (3)$$

and the perpendicular and parallel polarization vectors are given as

$$\vec{s}_{k_x,k_y} = \frac{\vec{k} \times \vec{z}}{\|\vec{k} \times \vec{z}\|},\ \vec{p}_{k_x,k_y} = \frac{\vec{s}_{k_x,k_y} \times \vec{k}}{\|\vec{k}\|}.\quad (4)$$

Using this decomposition, $\overrightarrow{E_{cam}}$ can be described as

$$\widetilde{\vec{E_{cam}}}(k_x,k_y) = \widetilde{E_s}(k_x,k_y)\vec{s}_{k_x,k_y} + \widetilde{E_p}(k_x,k_y)\vec{p}_{k_x,k_y}.\quad (5)$$

Here, $\widetilde{E_s}(k_x,k_y)$ and $\widetilde{E_p}(k_x,k_y)$ can be derived from the obtained fields $E_1$ and $E_2$. Using Eq. (1), the derivation is described as

$$\begin{pmatrix} \widetilde{E}_s(k_x,k_y) \\ \widetilde{E}_p(k_x,k_y) \end{pmatrix} = \begin{pmatrix} \vec{s}_{k_x,k_y} \cdot \vec{R_1} & \vec{p}_{k_x,k_y} \cdot \vec{R_1} \\ \vec{s}_{k_x,k_y} \cdot \vec{R_2} & \vec{p}_{k_x,k_y} \cdot \vec{R_2} \end{pmatrix}^{-1} \begin{pmatrix} \widetilde{E}_1(k_x,k_y) \\ \widetilde{E}_2(k_x,k_y) \end{pmatrix}. \quad (6)$$

The derived $\vec{E}_{cam}$ corresponds to the field on the camera plane. To describe the $\vec{E}_{sp}$ of the field on the sample plane, one should consider the effect of the system magnification ($M$) on the wave vector. Owing to the change in the wave vectors, the sample field can be described as

$$\widetilde{\vec{E}}_{sp}(Mk_x, Mk_y) = \widetilde{E}_s(k_x,k_y)\vec{s}_{k_x,k_y} + \widetilde{E}_p(k_x,k_y)\vec{p}_{Mk_x,Mk_y}. \quad (7)$$

Note that according to Eq. (4), the direction of p-pol is changed, but s-pol is not. Using Eq. (7), the plane waves of the sample field are individually vectorized in the Cartesian coordinate system.

The dielectric tensor tomogram was reconstructed from vectorial fields using the DTT algorithm [14]. For each illumination angle, three fields were acquired with different illumination polarizations and slightly tilted illumination.

$$\vec{F}(\vec{r}) = \begin{pmatrix} IFT^{3D}[-2ik_z e^{-ik_z z} \vec{p}_1 \circ \widetilde{\vec{\psi}_1}] \\ IFT^{3D}[-2ik_z e^{-ik_z z} \vec{p}_2 \circ \widetilde{\vec{\psi}_2}] \\ IFT^{3D}[-2ik_z e^{-ik_z z} \vec{p}_3 \circ \widetilde{\vec{\psi}_3}] \end{pmatrix} \begin{pmatrix} \vec{p}_1 \\ \vec{p}_2 \\ \{1 - i\vec{r} \cdot \Delta\vec{K}\}\vec{p}_3 \end{pmatrix}^{-1}, \quad (8)$$

where $\vec{F}$ denotes the dielectric tensor, $\vec{\psi}_i$ and $\vec{p}_i$ represent the scattered field and illumination polarization, with $i$ = 1, 2 denoting illuminations with right and left circular polarization and 3 denoting the differential quantities obtained by slightly tilting the illumination by $\Delta K$ [14]. After reconstruction, the dielectric tensors were diagonalized to obtain the principal RIs ($n_1$, $n_2$, and $n_3$) and their rotational transform ($R$) corresponding to the RI ellipsoid notation.

$$\overleftrightarrow{\varepsilon} = \begin{pmatrix} \varepsilon_{xx} & \varepsilon_{xy} & \varepsilon_{xz} \\ \varepsilon_{yx} & \varepsilon_{yy} & \varepsilon_{yz} \\ \varepsilon_{zx} & \varepsilon_{zy} & \varepsilon_{zz} \end{pmatrix} = R \begin{pmatrix} n_1^2 & 0 & 0 \\ 0 & n_2^2 & 0 \\ 0 & 0 & n_3^2 \end{pmatrix} R^T \quad (9)$$

Note that because the dielectric tensor is symmetric [9], six independent components of the dielectric tensor were reconstructed [Fig. 3(e)].

To validate the present method, we experimentally measured the 3D dielectric tensor tomograms of liquid-crystal (LC) particles (Figs. 4 and 5). These particles were prepared using a reactive mesogen mixture containing surfactants [26]. The surfactants, sodium dodecyl sulfate (SDS) and polyvinyl alcohol (PVA), were oriented in the direction of the optic axis with a radial and bipolar configuration.

The measured extraordinary [Figs. 4(a), (d)] and ordinary RIs [Figs. 4(b), (e)], as well as the optical axis directions [Figs. 4(c), (f)] are shown. Distributions agree with the expected radial and bipolar structures, and radial hedgehogs as well as boojum point defects were observed. Critical cross sections are also shown individually in Fig. 5, highlighting the point defects and their orientation configurations.

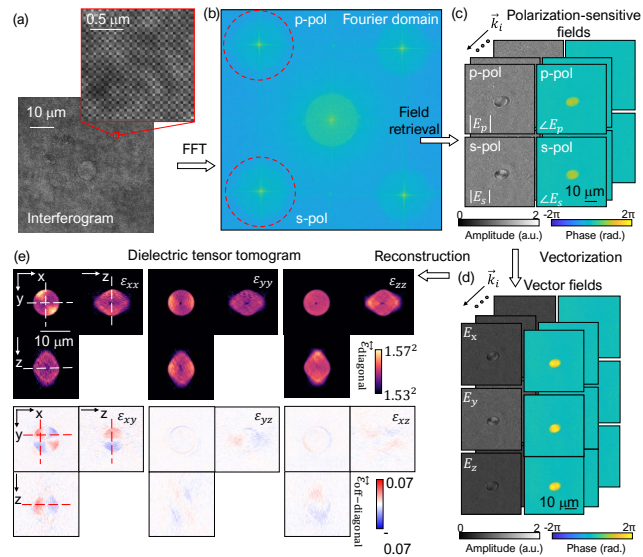

Fig. 3. Principle of DTT. (a) Multiplexed interferogram. Two interferograms are multiplexed as a check pattern. (b) Fourier transform of the interferogram. Components of the diffracted field parallel to the p-pol and s-pol reference beams are located at different areas in Fourier domain. (c) Retrieved fields. At various illumination wave vectors $k_i$, p-pol and s-pol fields are retrieved. (d) Vector fields at the sample plane. (e) Dielectric tensor tomogram reconstructed from data in (d) based on the DTT algorithm.

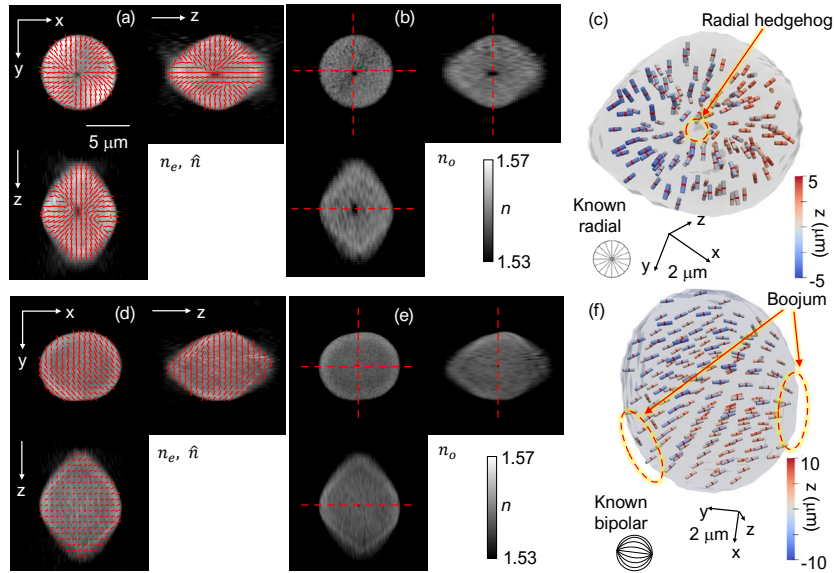

Fig. 4. Experimental validation of polarization multiplexed DTT. (a–c) Results of the radial LC particle. (d–f) Results of the bipolar LC particle. (a), (d) Cross-sections of extraordinary RIs $n_e$ mapped with their directions $\hat{n}$. (b), (e) Cross-sections of ordinary RIs $n_o$. The red dash lines indicate positions of the cross-sections. (c), (f) Directors $\hat{n}$ visualized in 3D. The isosurface of (c) is at $n = 1.5405$ and that of (f) is at $n = 1.5350$. The red dash circles indicate topological point defects of LC particles.

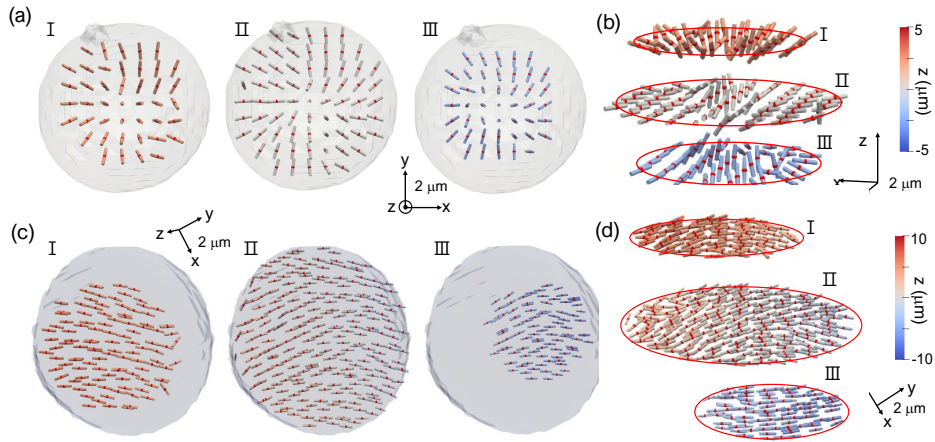

Fig. 5. Cross-sections of the 3D visualized directors $\hat{n}$. (a), (b) Cross-sections of figure 4(c). (c), (d) Cross-sections of figure 4(f). (a), (c) Top-view of the cross sections. (b), (d) Side-view of the cross sections. Roman numbers indicate the corresponding cross-sections.

In summary, we have presented mDTT, which uses only one camera to measure the anisotropy of microscopic objects. Compared with the original DTT method, no image registration is required in the proposed method, enabling simplification of the reconstruction code and improving the reconstruction accuracy.

At the camera, the two reference beams interfere with the sample beam, forming a multiplexed interferogram, from which the scattered vector field can be reconstructed. The dielectric tensor was retrieved from these vector fields by using the DTT algorithm. The dielectric tensor can be further analyzed to extract the anisotropy amplitude and direction to enable rich visualization of the anisotropic structure in liquid crystal particles. Using our method, an existing ODT setup based on angle scanning can be converted into a DTT setup.

Due to the present method avoid issues for field-image registration, it can be exploited for the precise high-resolution measurements of dielectric tensor distributions. For example, it can be used to investigate various underexplored applications where the quantification of 3D anisotropic structures and their orientational information are important, including anisotropic metamaterials [27, 28], liquid crystal materials [2, 29], collagen network in cancer tissues [30, 31], and anisotropic structures in sperm cells or oocytes [32-34].

**Funding.** This work was supported by KAIST UP program, BK21+ program, Tomocube Inc., National Research Foundation of Korea (2015R1A3A2066550), KAIST Institute of Technology Value Creation, Industry Liaison Center (G-CORE Project) grant funded by the Ministry of Science and ICT (N11210014, N11220131) and Institute of Information & Communications Technology Planning & Evaluation (IITP; 2021-0-00745) grant funded by the Korea government (MSIT).

**Disclosures.** The authors declare that there are no conflicts of interest related to this article.

**Data availability.** Data underlying the results presented in this paper are not publicly available at this time but may be obtained from the authors upon reasonable request.

## References

1. D. Needleman, and Z. Dogic, Nature reviews materials **2**, 1-14 (2017).
2. S. J. Woltman, G. D. Jay, and G. P. Crawford, Nature materials **6**, 929-938 (2007).
3. S. L. Jacques, Physics in Medicine & Biology **58**, R37 (2013).
4. M. Koike‐Tani, T. Tani, S. B. Mehta, A. Verma, and R. Oldenbourg, Molecular reproduction and development **82**, 548-562 (2015).
5. R. Oldenbourg, Cold Spring Harbor Protocols **2013**, pdb. top078600 (2013).
6. B. Ge, Q. Zhang, R. Zhang, J.-T. Lin, P.-H. Tseng, C.-W. Chang, C.-Y. Dong, R. Zhou, Z. Yaqoob, and I. Bischofberger, ACS Photonics **8**, 3440-3447 (2021).
7. Y. Kim, J. Jeong, J. Jang, M. W. Kim, and Y. Park, Optics Express **20**, 9948-9955 (2012).
8. T. Colomb, F. Dürr, E. Cuche, P. Marquet, H. G. Limberger, R.-P. Salathé, and C. Depeursinge, Applied optics **44**, 4461-4469 (2005).
9. M. Born, and E. Wolf, *Principles of optics: Electromagnetic theory of propagation, interference and diffraction of light* (Elsevier, 2013).
10. Y. Park, C. Depeursinge, and G. Popescu, Nature photonics **12**, 578-589 (2018).
11. O. Haeberlé, K. Belkebir, H. Giovaninni, and A. Sentenac, Journal of Modern Optics **57**, 686-699 (2010).
12. A. Saba, J. Lim, A. B. Ayoub, E. E. Antoine, and D. Psaltis, Optica **8**, 402-408 (2021).
13. J. van Rooij, and J. Kalkman, Biomed. Opt. Express **11**, 2109-2121 (2020).
14. S. Shin, J. Eun, S. S. Lee, C. Lee, H. Hugonnet, D. K. Yoon, S.-H. Kim, J. Jeong, and Y. Park, Nature Materials **21**, 317-324 (2022).
15. J. Jung, J. Kim, M.-K. Seo, and Y. Park, Optics Express **26**, 7701-7711 (2018).
16. T. D. Yang, K. Park, Y. G. Kang, K. J. Lee, B.-M. Kim, and Y. Choi, Optics Express **24**, 29302-29311 (2016).
17. K. Park, T. D. Yang, D. Seo, M. G. Hyeon, T. Kong, B.-M. Kim, Y. Choi, W. Choi, and Y. Choi, ACS Photonics **8**, 3042-3050 (2021).
18. A. W. Lohmann, Appl. Opt. **4**, 1667-1668 (1965).
19. X. Liu, B.-Y. Wang, and C.-S. Guo, Opt. Lett. **39**, 6170-6173 (2014).
20. N. T. Shaked, V. Micó, M. Trusiak, A. Kuś, and S. K. Mirsky, Adv. Opt. Photon. **12**, 556-611 (2020).
21. S. Shin, K. Kim, J. Yoon, and Y. Park, Opt. Lett. **40**, 5407-5410 (2015).
22. K. Lee, K. Kim, G. Kim, S. Shin, and Y. Park, Opt. Lett. **42**, 999-1002 (2017).
23. M. Takeda, H. Ina, and S. Kobayashi, JosA **72**, 156-160 (1982).
24. S. K. Debnath, and Y. Park, Opt. Lett. **36**, 4677-4679 (2011).
25. K. Kim, J. Yoon, S. Shin, S. Lee, S.-A. Yang, and Y. Park, Journal of Biomedical Photonics & Engineering **2**, 020201-020201 (2016).
26. J.-H. Lee, T. Kamal, S. V. Roth, P. Zhang, and S.-Y. Park, RSC Advances **4**, 40617-40625 (2014).
27. J. Hao, Y. Yuan, L. Ran, T. Jiang, J. A. Kong, C. Chan, and L. Zhou, Physical review letters **99**, 063908 (2007).
28. L. H. Nicholls, F. J. Rodríguez-Fortuño, M. E. Nasir, R. M. Córdova-Castro, N. Olivier, G. A. Wurtz, and A. V. Zayats, Nature Photonics **11**, 628-633 (2017).
29. M. Schadt, Annual review of materials science **27**, 305-379 (1997).
30. M. D. Shoulders, and R. T. Raines, Annual review of biochemistry **78**, 929 (2009).
31. K. M. Riching, B. L. Cox, M. R. Salick, C. Pehlke, A. S. Riching, S. M. Ponik, B. R. Bass, W. C. Crone, Y. Jiang, and A. M. Weaver, Biophysical journal **107**, 2546-2558 (2014).
32. L. Gianaroli, M. C. Magli, A. P. Ferraretti, A. Crippa, M. Lappi, S. Capitani, and B. Baccetti, Fertility and sterility **93**, 807-813 (2010).
33. W. Wang, L. Meng, R. Hackett, and D. Keefe, Human Reproduction **16**, 1464-1468 (2001).
34. C. Madaschi, T. Aoki, D. P. d. A. F. Braga, R. d. C. S. Figueira, L. S. Francisco, A. Iaconelli Jr, and E. Borges Jr, Reproductive biomedicine online **18**, 681-686 (2009).